\RequirePackage{fix-cm}
\documentclass{svjour3}                     
\smartqed  
\usepackage{graphicx}
\usepackage{amsmath}
\usepackage{algorithm}
\usepackage{algpseudocode}
\usepackage{algorithmicx}
\usepackage{hyperref}
\usepackage{tabularx}
\usepackage{color}
\begin{document}


\title{Contextualization of topics}

\subtitle{Browsing through the universe of bibliographic information}

\author{Rob Koopman         \and
        Shenghui Wang \and 
        Andrea Scharnhorst
}

\institute{R. Koopman \at
              OCLC Research, Schipholweg 99, Leiden, The Netherlands \\
              Tel.: +31 71 524 6500\\
              \email{rob.koopman@oclc.org}           
           \and
           S. Wang \at
              OCLC Research, Schipholweg 99, Leiden, The Netherlands \\
              Tel.: +31 71 524 6500\\
              \email{shenghui.wang@oclc.org}           
			\and 
            A. Scharnhorst \at
              DANS-KNAW, Anna van Saksenlaan 51, The Hague, The Netherlands \\
              Tel.: +31 70 349 4450\\
              \email{andrea.scharnhorst@dans.knaw.nl}           
}

\date{Received: date / Accepted: date}

\maketitle

\begin{abstract}
This paper describes how semantic indexing can help to generate a contextual overview of topics and visually compare clusters of articles. The method was originally developed for an innovative information exploration tool, called \textit{Ariadne}, which operates on bibliographic databases with tens of millions of records~\cite{koopman2015_chi}. In this paper, the method behind \textit{Ariadne} is further developed and applied to the research question of the special issue ``Same data, different results" -- the better understanding of topic (re-)construction by different bibliometric approaches. For the case of the \textit{Astro} dataset of 111,616 articles in astronomy and astrophysics, a new instantiation of the interactive exploring tool, \textit{LittleAriadne}, has been created. This paper contributes to the overall challenge to delineate and define topics in two different ways. First, we produce two clustering solutions based on vector representations of articles in a lexical space. These vectors are built on semantic indexing of entities associated with those articles. Second, we discuss how \textit{LittleAriadne} can be used to browse through the network of topical terms, authors, journals, citations and various cluster solutions of the \textit{Astro} dataset. More specifically, we treat the assignment of an article to the different clustering solutions as an additional element of its bibliographic record. Keeping the principle of semantic indexing on the level of such an extended list of entities of the bibliographic record, \textit{LittleAriadne} in turn provides a visualization of the context of a specific clustering solution. It also conveys the similarity of article clusters produced by different algorithms, hence representing a complementary approach to other possible means of comparison. 

\end{abstract}

\section{Introduction}
\label{sec.intro}

What is the essence, or the boundary of a scientific field? How can a topic be defined? Those questions are at the heart of bibliometrics. They are equally relevant for indexing, cataloguing and consequently information retrieval~\cite{Mayr2015}. Rigour and stability in  bibliometrically defining boundaries of a field are important for research evaluation and consequently the distribution of funding. But, for information retrieval - next to accuracy - serendipity, broad coverage and associations to other fields are of equal importance. If researchers seek information about a certain topic outside of their areas of expertise, their information needs can be quite different from those in a bibliometric context. Among the many possible hits for a search query, they may want to know which are core works (articles, books) and which are rather peripheral. They may want to use different rankings~\cite{Mutschke2014}, get some additional context information about authors or journals, or see other closely related vocabulary or works associated with a search term. On the whole, they would have less need to define a topic and a field in a bijective, univocal way. Such a possibility to contextualize is not only important for term-based queries. It also holds for groups of query terms, or for the exploration of sets of documents, produced by different clustering algorithms. Contextualisation is the main motivation behind this paper.

If we talk of contextualisation we still stay in the realm of bibliographic information. That is, we rely on information about authors, journals, words, references as hidden in the entirety of the set of all bibliographic records. Decades of bibliometrics research have produced many different approaches to cluster documents, or more specifically, articles. They often focus on one entity of the bibliographic record. To give one example, articles and terms within those articles (in title, abstract and/or full text) form a bipartite network. From this network we can either build a network of related terms (co-word analysis) or a network of related articles (based on shared words). The first method, sometimes  called lexical~\cite{leydesdorff2011a}, has been applied in scientometrics to produce so-called topical or semantic maps. The same exercise can be applied to authors and articles, authors and words~\cite{lu2012}, and in effect to each element of the bibliographic record for an article~\cite{havemann2012}. If we extend the bibliographic record of an article with the list of references contained by this article, we enter the area of citation analysis. Here, the following methods are widely used: direct citations, bibliographic coupling 
and co-citation maps. 
Hybrid methods combine citation and lexical analysis (e.g.,~\cite{Janssens2009,Zitt2006}).  We would like to note here that in an earlier comparison of citation- and word-based mapping approaches Zitt et al. (\cite{Zitt2011}) underline the differences both signals carry in terms of what aspect of scientific practice they represent. We come back to this in the next paragraph.  
Formally spoken, the majority of studies apply one method and often display unipartite networks. Sometimes analysis and visualization of multi-partite networks can be found~\cite{heur2013}. 

Each network representation of articles captures some aspect of connectivity and structure which can be found in published work. Co-authorship networks shed light on the social dimension of knowledge production, the so-called Invisible College~\cite{glanzel2004,mali2012}. Citation relations are interpreted as traces of flows of knowledge~\cite{radicchi2012,price1965}. By using different bibliographic elements, we obtain different models for, or representations of, a field or topic; i.e. as a conceptual, cognitive unit; as a community of practice; or as institutionalized in journals. One could also say that choosing \textit{what to measure} affects the representation of a field or topic. Another source of variety beyond differences arising from choice of representations is how to analyze those representations. Fortunately, network analysis provides several classical methods to choose from, including clustering and clique analysis. However, clusters can be defined in different ways, and some clustering algorithms can be computationally expensive when used on large or complex networks. Consequently, we find different solutions for the same algorithm (if parameters in the algorithm are changed) and different solutions for different algorithms. One could call this an effect of the choice of instrument for the measurement or \textit{how to measure}. Using an ideal-typical workflow, these points of choice have been further detailed and discussed in another paper of this special issue (~\cite{velden2015comparison}). The variability in each of the stages of the workflow results in ambiguity, and, if not  articulated, makes it even harder to reproduce results. Overall, moments of choice add an uncertainty margin to the results~\cite{Kouw2013,Petersen2006}. Last but not least, we can ask ourselves whether clear delineations exist between topics in practice. Often in the sciences very different topics are still related to each other. There exist unsharp boundaries and almost invisible long threads in the fabric of science~\cite{boyack2010}, which might inhibit the finding of a contradiction-free solution in form of a unique set of disjunct clusters. There is a seeming paradox between the fact that experts often can rather clearly identify what belongs to their field or a certain topic, and that it is so hard to quantitatively represent this with bibliometric methods. However, a closer look into science history and science and technology studies reveals that even among experts opions regarding subject matter classification or topic identification might vary. What belongs to a field and what not is as much an epistemic question as also an object of social negotiations. Moreover, the boundaries of a field change over time, and even a defined canon or body of knowledge determining the essence of a field or a topic can still be controversial or subject to change~\cite{galison1997image}. 

Defining a topic requires a trade-off between accepting the natural ambiguity of what a topic is and the necessity to define a topic for purposes of education, knowledge acquisition, and evaluation. Since different perspectives serve different purposes, there is also a need to preserve the diversity and ambiguity described earlier. Having said this, for the sake of scientific reasoning it is equally necessary to be able to further specify the validity and appropriateness of different methods for defining topics and fields~\cite{intro}.

This paper contributes to this sorting-out-process in several ways. All are driven by the motivation to provide a better understanding of the topic re-construction results by providing context: context of the topics themselves by using a lexical approach and \textit{all} elements of the bibliographical record to delineate topics;  and context for different solutions in the (re-)construction of topics. We first introduce the method of semantic indexing, by which each bibliographic record is decomposed and a vector representation for each of its entities in a lexical space is build, resulting in a so-called \textit{semantic matrix}. This approach is conceptually closer to classical information retrieval techniques based on Salton's vector space model~\cite{salton1986} than to the usual bibliometrical mapping techniques. In particular, it is similar to Latent Semantic Indexing or Latent Semantic Analysis. In the specific case of the \textit{Astro} dataset, we extend the bibliographic record with information on cluster assignments provided by different clustering solutions. For the purpose of a delineation of topics based on clustering of articles, we reconstruct a semantic matrix for articles based on the semantic indexing of their individual entities. Secondly, based on this second matrix, we produce our own clustering solutions (detailed in \cite{koopman2015_clustering}) by applying two different clustering algorithms.  Third, we present an interactive visual interface called \textit{LittleAriadne} that displays the context around those extracted entities. The interface responds to a search query with a network visualization of most related terms, authors, journals, citations and cluster IDs. The query can consist of words or author names, but also clustering solutions. The displayed nodes or entities around a query term represent, to a certain extent, the context of the query in a lexical, semantic space. 

In what follows, we address the following research questions:  
\begin{description}
\item[Q1] How does the \textit{Ariadne} algorithm, originally developed for a large corpora which contains tens of millions of articles, work on a much smaller, field-specific dataset?  How can we relate the produced contexts to domain knowledge retrieved from other information services?
\item[Q2] Can we use \textit{LittleAriadne} to compare different cluster assignments of papers, by treating those cluster assignments as additional entities? What can we learn about the topical nature of these clusters when exploring them visually?
\end{description}
Concerning the last question, we restrict this paper to a description of the approach \textit{LittleAriadne} offers, and we provide some illustrations. A more detailed discussion of the results of this comparison has been taken up as part of the comparison paper of this special issue \cite{velden2015comparison}, which on the whole addresses different analytic methods and visual means to compare different clustering solutions.

\section{Data}
The \textit{Astro} dataset used in this paper contains documents published in the period 2003--2010 in 59 astrophysical journals.\footnote{For details of the data collection and cleaning process leading to the common used \textit{Astro} dataset see \cite{velden2015comparison}.}
Originally, these documents had been downloaded from the Web of Science in the context of a German-funded research project called ``Measuring Diversity of Research,'' conducted at the Humboldt-University Berlin from 2009 to 2012. Based on institutional access to the Web of Science, we worked on the same dataset. Starting with 120,007 records in total, 111,616 records of the document types Article, Letter and Proceedings Paper have been treated with different clustering methods (see the other contributions to this special issue). 

Different clustering solutions have been shared, and eventually a selection of solutions for the comparison has been defined. In our paper we used clustering solutions from CWTS-C5 (c)~\cite{VanEck2017}, UMSI0 (u)~\cite{Velden2017}, HU-DC (hd)~\cite{Havemann2017}, STS-RG (sr)~\cite{Boyack2017}, ECOOM-BC13 (eb), ECOOM-NLP11 (en) (both ~\cite{Glaenzel2017}) and two of our own: OCLC-31 (ok) and OCLC-Louvain (ol)~\cite{koopman2015_clustering}.
The CWTS-C5 and UMSI0 are the clustering solutions generated by two different methods, Infomap and the Smart Local Moving Algorithm (SLMA) respectively, applied on the same direct citation network of articles. The two ECOOM clustering solutions are generated by applying the Louvain method to find communities among bibliographic coupled articles where ECOOM-NLP11 also incorporates the keywords information. The STS-RG clusters are generated by first projecting the relatively small \textit{Astro} dataset to the full Scopus database. After the full Scopus articles are clustered using SLMA 
on the direct citation network, the  cluster assignments of \textit{Astro} articles are collected. The HU-DC clusters are the only overlapping clusters generated by a memetic type algorithm designed for the extraction of overlapping, poly-hierarchical topics in the scientific literature. 
Each article is assigned to a HU-DC cluster with a confidence value. We only took those assignments with a confidence value higher than 0.5. More detailed accounts of these clustering solutions can be found in~\cite{velden2015comparison}.
  Table~\ref{tab.clusters} shows their labels later used in the interface, and how many clusters each solution produced. All the clustering solutions are based on the full dataset. However, each article is not necessarily guaranteed to have a cluster assignment in every clustering solution (see the papers about the clustering solutions for further details). The last column in Table~\ref{tab.clusters} shows how many articles of the original dataset are covered by different solutions.

\begin{table}[ht]
\centering
\caption{ Statistics of clustering solutions generated by different methods}
\label{tab.clusters}
\begin{tabular}{cccc}\\\hline
Cluster label & Solution & \#Clusters & Coverage\\\hline
c & CWTS-C5 	& 22	& 91\%\\
u & UMSI0 	& 22	& 91\%\\
ok & OCLC-31 & 31	& 100\%\\
ol & OCLC-Louvain & 32	& 100\%\\
sr & STS-RG		& 556 & 96\%\\ 
eb & ECOOM-BC13	& 13	& 97\%\\
en & ECOOM-NLP11	& 11 & 98\%\\
hd & HU-DC	& 113	& 91\%\\\hline
\end{tabular}
\end{table}

\section{Method}

\subsection{Building semantic representations for entities}
\label{sec.rp}
The \textit{Ariadne} algorithm was originally developed on top of the article database, ArticleFirst of OCLC~\cite{koopman2015_chi}. The interface, accessible at \url{http://thoth.pica.nl/relate}, allows users to visually and interactively browse through 35 thousand journals, 3 million authors, and 1 million topical terms associated with 65 million articles. The \textit{Ariadne} pipeline consists of two steps: an offline procedure for semantic indexing and an online interactive visualization of the context of search queries. We applied the same method to the \textit{Astro} dataset and built an instantiation, named \textit{LittleAriadne}, accessible at \url{http://thoth.pica.nl/astro/relate}. 

\begin{table}[h]
\caption{An article from the \textit{Astro} dataset}
\label{tab.article}
\footnotesize { 
\begin{tabularx}{\textwidth}{lX}\hline
Article ID & ISI:000276828000006 \\\hline
Title & On the Mass Transfer Rate in SS Cyg\\\hline
Abstract & 
The mass transfer rate in SS Cyg at quiescence, estimated 
from the observed luminosity of the hot spot, is log M-tr 
= 16.8 +/- 0.3. This is safely below the critical mass 
transfer rates of log M-crit = 18.1 (corresponding to log 
T-crit(0) = 3.88) or log M-crit = 17.2 (corresponding to 
the ``revised'' value of log T-crit(0) = 3.65). The mass 
transfer rate during outbursts is strongly enhanced\\\hline
Author & [author:smak j]\\\hline
ISSN & [issn:0001-5237]\\\hline
Subject 
& [subject:accretion, accretion disks] [subject:cataclysmic 
variables] [subject:disc instability model] [subject:dwarf novae] 
[subject:novae, cataclysmic variables] [subject:outbursts] 
[subject:parameters] [subject:stars] [subject:stars dwarf novae] 
[subject:stars individual ss cyg] [subject:state] [subject:
superoutbursts] \\\hline
Citation & [citation:bitner ma, 2007, astrophys j 1, v662, p564] [citation:bruch a, 1994, astron astrophys sup, v104, p79] [citation:buatmenard v, 2001, astron astrophys, v369, p925] [citation:hameury jm, 1998, mon not r astron soc, v298, p1048] [citation:harrison te, 1999, astrophys j 2, v515, l93] [citation:kjuikchieva d, 1998, a as, v262, p53] [citation:kraft rp, 1969, apj, v158, p589] [citation:kurucz rl, 1993, cd rom] [citation:lasota jp, 2001, new astron rev, v45, p449] [citation:paczynski b, 1980, acta astron, v30, p127] [citation:schreiber mr, 2002, astron astrophys, v382, p124] [citation:schreiber mr, 2007, astron astrophys, v473, p897, doi 10.1051/0004-6361:20078146] [citation:smak j, 1996, acta astronom, v46, p377] [citation:smak j, 2002, acta astronom, v52, p429] [citation:smak j, 2004, acta astronom, v54, p221] [citation:smak j, 2008, acta astronom, v58, p55] [citation:smak ji, 2001, acta astronom, v51, p279] [citation:tutukov av, 1985, pisma astron zh, v11, p123] [citation:tutukov av, 1985, sov astron lett+, v11, p52] [citation:voloshina ib, 2000, astron rep+, v44, p89] [citation:voloshina ib, 2000, astron zh, v77, p109] 	\\\hline
Topical terms 
& mass transfer; transfer rate; ss; cyg; quiescence; estimated; observed; luminosity; hot spot; log; tr; safely; critical; crit; corresponding; revised; value; outbursts; strongly; enhanced \\\hline
UAT terms&
[uat:stellar phenomena]; [uat:mass transfer]; [uat:optical bursts]\\\hline
Cluster ID& [cluster:c 19] [cluster:u 16] [cluster:ok 18] [cluster:ol 23] [cluster:sr 17] [cluster:eb 1] [cluster:en 1] [cluster:hd 1] [cluster:hd 18] [cluster:hd 48]\\\hline
\end{tabularx}
}
\end{table}

To describe our method we give  an example of an article from the \textit{Astro} dataset in table~\ref{tab.article}. We list all the fields of this bibliographic record that we used for \textit{LittleAriadne}. We include the following types of entities for semantic indexing: authors, journals (ISSN), subjects, citations, topical terms, MAI-UAT thesaurus terms and cluster IDs (see Table~\ref{tab.clusters}). For the \textit{Astro} dataset, we extended the original \textit{Ariadne} algorithm~\cite{koopman2015_issi} by adding citations as additional entities. 
In the short paper about the OCLC clustering solutions~\cite{koopman2015_clustering} we applied clustering to different variants of the vector representation of articles, including variants with and without citations. We reported there about the effect of adding citations to vector representations of articles on clustering. 

In Table~\ref{tab.article} we display the author name (and other entities) in a syntax (indicated by square brackets) that can immediately be used in the search field of the interface. Each author name is treated as a separate entity. The next type of entity is the journal identified by its ISSN number. One can search for a single journal using its ISSN number. In the visual interface, the ISSN numbers are replaced by the journal name, which is used as label for a journal node. The next type of entities are so-called subjects. Those subjects originate from the fields ``Author Keywords'' and ``Keywords Plus'' of the original Web of Science records. Citations, references in the article, are considered as a type of entity too. Here, we use the standardized abbreviated citations in the Web of Science database. We remark that we do not apply any form of disambiguation--neither for the author names nor for the citations. Topical terms such as ``mass transfer'' and ``quiescence'' in our example, are single words or two-word phrases extracted from titles and abstracts of all documents in the dataset. A multi-lingual stop-word list was used to remove unimportant words, and mutual information was used to generate two-word phrases. Only words and phrases which occur more than a certain threshold value were kept.  

The next type of entity is a set of Unified Astronomy Thesaurus (UAT)\footnote{\url{http://astrothesaurus.org/}} terms which were assigned by the Data Harmony's Machine Aided Indexer (M.A.I.).\footnote{\url{http://www.dataharmony.com/services-view/mai/}} Please refer to~\cite{kevin2015_thesaurus} for more details about the thesaurus and the indexing procedure. 
The last type of entity we add to each of the articles (specific for \textit{LittleAriadne}) is the collection of cluster IDs corresponding to the clusters to which the article was assigned by the various clustering algorithms. For example, the article in Table~\ref{tab.article} has been assigned to clusters ``c 19'' (produced by CWTS-C5) and  ``u 16'' (produced by UMSI0), and so on. In other words, we treat the cluster assignments of articles as they would be classification numbers or additional subject headings. Table~\ref{tab.entities} lists the total number of different types of entities found in the \textit{Astro} dataset.
\begin{table}[t]
\centering
\caption{Entities in \textit{LittleAriadne}}
\label{tab.entities}
\begin{tabular}{lr}\\
Journals & 59 \\
Authors & 55,607 \\
Topical terms & 60,501 \\
Subjects & 41,945 \\
Citations & 386,217 \\
UAT terms & 1534 \\
Cluster IDs & 610 \\\hline
Total & 546,473\\
\end{tabular}
\end{table}

\begin{figure}
\centering
\includegraphics[width=\linewidth]{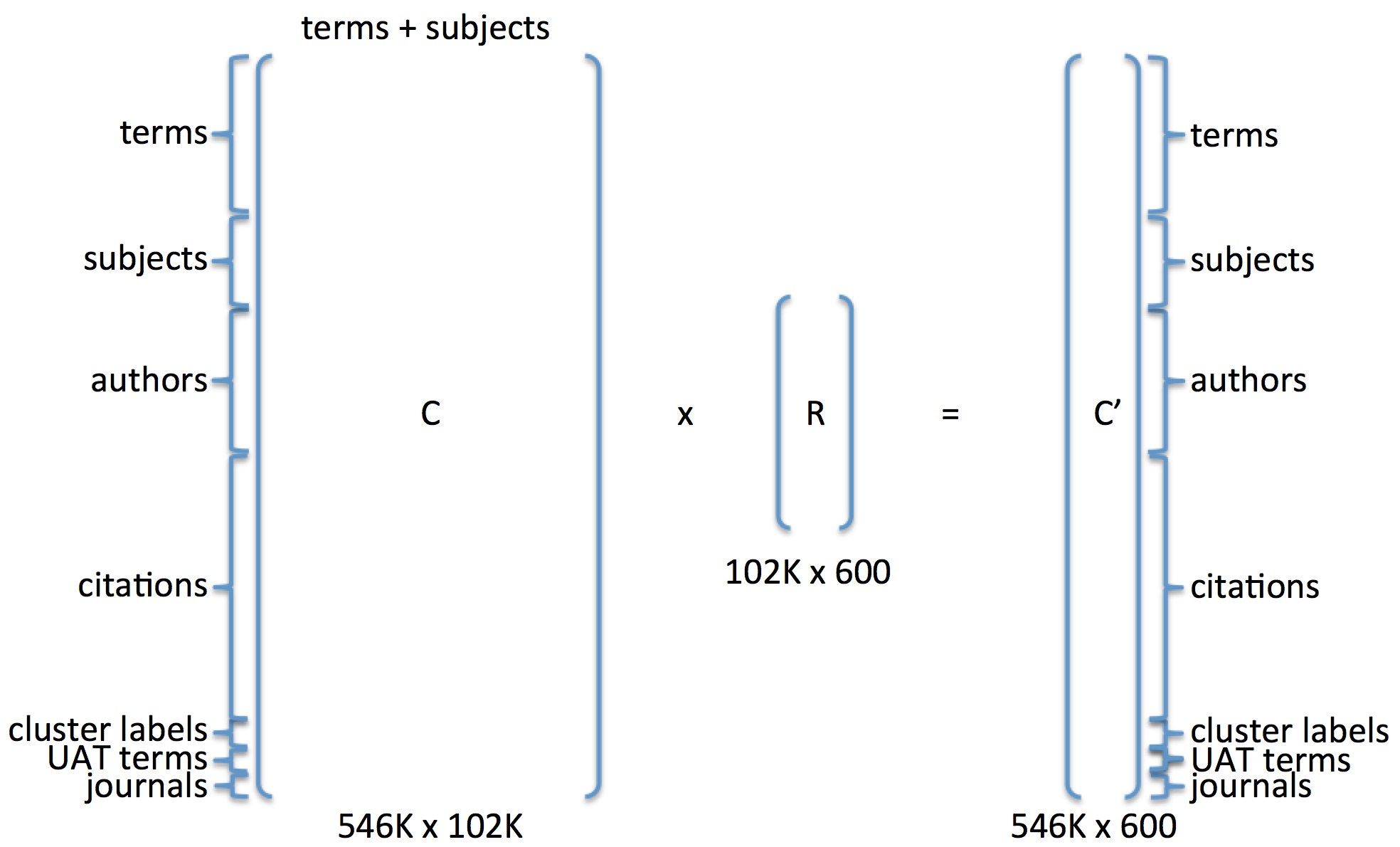}
\caption{Dimension reduction using Random Projection \label{fig.rp}}
\end{figure}

To summarize, we deconstruct each bibliographic record, extract a number of entities, and add some more (the cluster IDs and the topical terms). Next, we construct for each of these entities a vector in a word space built from topical terms and subject terms. We assume that the context of all entities is captured by their vectors in this space. Figure~\ref{fig.rp} gives a schematic representation of these vectors which form the matrix $C$. All types of entities -- topical term, subject, author, citation, cluster ID and journal -- form the rows of the matrix, and their components (all topical terms and subjects) the columns. The values of the vector components are the frequencies of the co-occurrence of an entity and a specific word in the whole dataset. That is, we count how many articles contain both an entity and a certain topical term or subject. 

Matrix $C$ expresses the semantics of all entities in terms of their context. Such context is then used in a computation of their similarity/relateness. Each vector can be seen as the lexical profile of a particular entity. A high cosine similarity value between two entities indicates a large overlap of the contexts of these two entities -- in other words, a high similarity between them. This is different from measuring their direct co-occurrence. 

For \textit{LittleAriadne}, the matrix $C$ has roughly $546K \times 102K$ elements, and is sparse and expensive for computation. To make the algorithm scale and to produce a responsive online visual interface, we applied the method of Random Projection~\cite{Achlioptas2003671,johnson84extensionslipschitz} to reduce the dimensionality of the matrix. As shown in Figure~\ref{fig.rp}, we multiply $C$ with a $102K \times 600$ matrix of randomly distributed --1 and 1, with half-half probabilities.\footnote{More efficient random projections are available. This version is more conservative and also computationally easier.} This way, the original $546K \times 102K$ matrix $C$ is reduced to a Semantic Matrix $C'$ of the size of $546K \times 600$. Still, each row vector represents the semantics of an entity. It has been discussed elsewhere~\cite{Bingham2001} that with the method of Random Projection, similar to other dimension reduction methods, essential properties of the original vector space are preserved, and thus entities with a similar profile in the high-dimensional space  still have a similar profile in the reduced space. A big advantage of Random Projection is that the computation is significantly less expensive than other methods, e.g., Principal Component Analysis~\cite{Bingham2001}. Actually, Random Projection is often suggested as a way of speeding up Latent Semantic Indexing (LSI)~\cite{Papadimitriou2000}, and \textit{Ariadne} is similar to LSI in some ways. LSI starts from a weighted term-document matrix, where each row represents the lexical profile of a document in a word space. In \textit{Ariadne}, however, the unit of analysis is not the document. Instead, each entity of the bibliographic record is subject to a lexical profile. We explain in the next section that, by aggregating over all entities belonging to one article, one can construct a vector representation for the article that represents its semantics and is suitable for further clustering processes (for more details please consult~\cite{koopman2015_clustering}).

With the Matrix $C'$, the interactive visual interface dynamically computes the most related entities (i.e., ranked by cosine similarity) to a search query. After irrelevant entities have been filtered out by removing entities with a high Mahalanobis distance~\cite{mahalanobis1936} to the query, the remaining entities and the query node are positioned in 2D so that the distance between nodes preserves the corresponding distance in the high dimensional space as much as possible. We use a spring-like force-directed graph drawing algorithm for the positioning of the nodes. Designed as experimental, explorative tool, no other optimisation of the network layout is applied. In the on-line interface, it is possible to zoom into the visualization, to change the size of the labels (\textit{font} slider) as well as the number of entities displayed (\textit{show} slider). For the figures in the paper, we used snapshots, in which node labels might overlap. Therefore, we provide links to the corresponding interactive display for each of the figures.   
In the end, with its most related entities, the context of a query term can be effectively presented~\cite{koopman2015_chi}. 
For \textit{LittleAriadne} we extended the usual \textit{Ariadne} interface with different lists of the most related entities, organized by type. This information is given below the network visualization.


\subsection{From a semantic matrix of entities to a semantic matrix for articles}
\label{sec.weightedavg}


The \textit{Ariadne} interface provides context around entities, but  does not produce article clusters directly. In other words, articles contribute to the context of entities associated with them but the semantics of themselves need to be reconstructed before we can apply clustering methods to identify article clusters. We describe the OCLC clustering workflow elsewhere~\cite{koopman2015_clustering}, but here we would like to explain the preparatory work for it. 


The first step is to create a vector representation of each article. For each article, we look up all entities associated with this article in the Semantic Matrix $C'$. We purposefully leave out the cluster IDs, because we want to construct our own clustering later independently, i.e., without already including information about clustering solutions of other teams.  For each article we obtain a set of vectors. 
For our article example in Table~\ref{tab.article} we have 55 entities. The set of vectors for this article entails one vector representing the single author of this article, 12 vectors for the subjects, one vector for the journal, 21 vectors for the citations and 20 vectors for topical terms. 
Each article is represented by a unique set of vectors. The size of the set can vary, but each of the vectors inside of a set has the same length, namely 600. 

For each article we compute the weighted average of its constituent vectors as its semantic representation.  
Each entity is weighted by its inverse document frequency to the third power; therefore, frequent entities are heavily penalized to have little contribution to the resulting representation of the article. In the end, each article is represented by a vector of 600 dimensions which becomes a row in a new matrix $M$ with the size of $111,616 \times 600$. Note that since articles are represented as a vector in the same space where other entities are also represented, it is now possible to compute the relatedness between entities and articles! Therefore in the online interface, we can present the articles most related to a query.

To group these 111,616 articles into meaningful clusters, we apply standard clustering methods to $M$. A first choice, the K-Means clustering algorithm results in 31 clusters. As detailed in~\cite{koopman2015_clustering}, with $k=31$, the resulting 31 clusters perform the best according to a pseudo-ground-truth built from the consensus of CWTS-C5, UMSI0, STS-RG and ECOOM-BC13. With this clustering solution the whole dataset is partitioned pretty evenly: the average size is 3600 $\pm$ 1371, and the largest cluster contains 6292 articles and the smallest 1627 articles. 

We also apply a network-based clustering method: the Louvain community detection algorithm. 
To avoid high computational cost, we first calculate for each article the top 40 most related articles, i.e., with the highest cosine similarity. This results in a new adjacency matrix $M'$ between articles, representing an article similarity network where the nodes are articles and the links indicate that the connected articles are very similar. We set the threshold for the cosine similarity at 0.6 to reduce links with low similarity values. A standard Louvain community detection algorithm~\cite{louvain} is applied to this network, producing 32 partitions, i.e., 32 clusters. 
Compared to K-Means 31 clusters, these 32 Louvain clusters vary more in terms of cluster size, with the largest cluster containing 9464 articles while the smallest cluster 86 articles. The Normalized Mutual Information~\cite{nami} between these two solutions is 0.68, indicating that they are highly similar to each other yet different enough to be studied further. More details can be found in~\cite{koopman2015_clustering}. 

\section{Experiments and results}

To answer the two research questions listed in the introduction, we conducted the following experiments: 
\begin{description}
\item [Experiment 1.] We implemented \textit{LittleAriadne} as an information retrieval tool. We searched with query terms, inspected and navigated through the resulting network visualization.
\item [Experiment 2.] We visually observed and compared different clustering solutions.
\end{description}

\subsection{Experiment 1 -- Navigate through networked information}

\begin{figure}
\includegraphics[width=\linewidth]{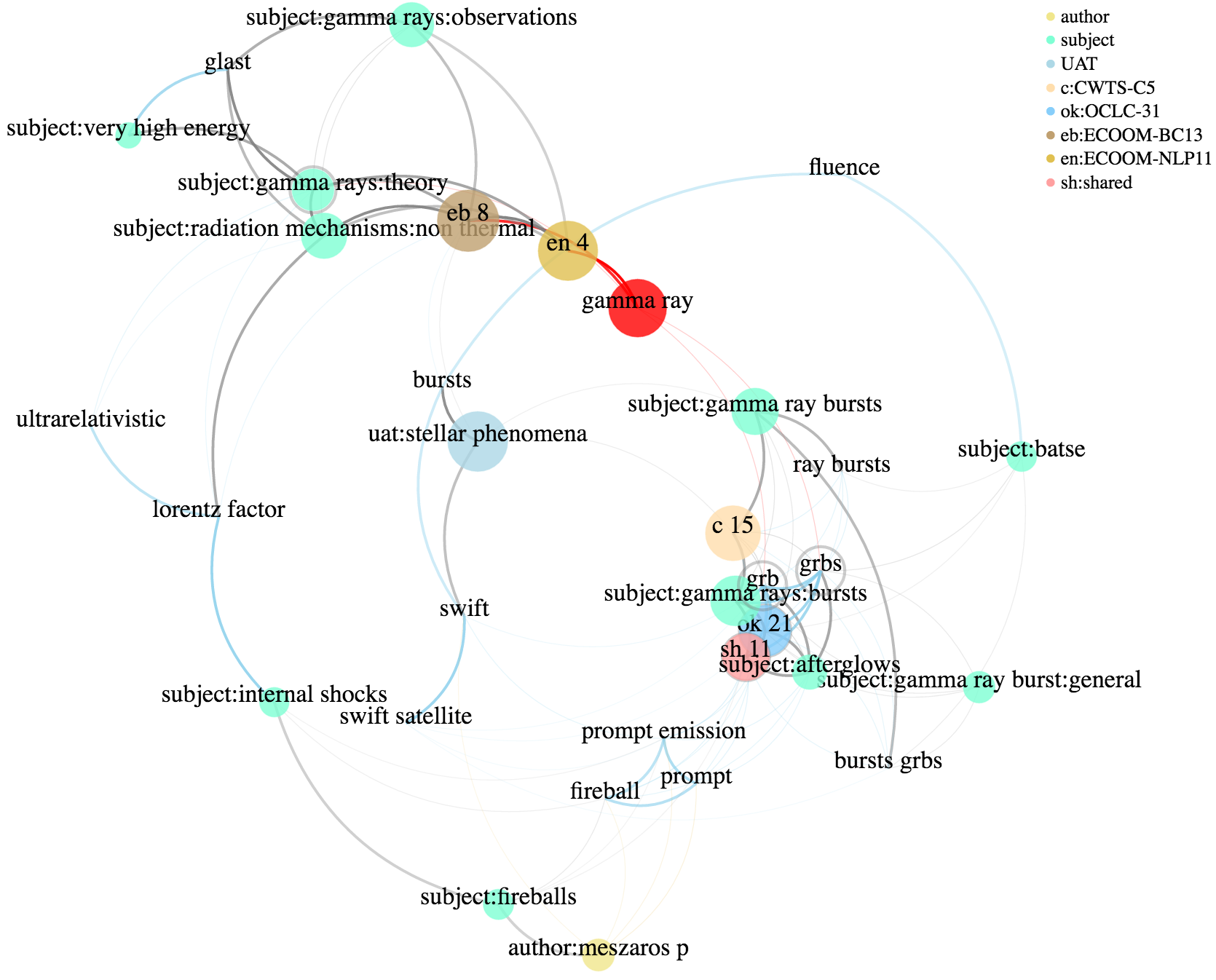}
\caption{The contextual view of the query term ``gamma ray''\label{fig.gammaray}}
\end{figure}
We implemented \textit{LittleAriadne}, which allows users to browse the context of the 546K entities associated with 111K articles in the datasets. If the search query refers to an entity that exists in the semantic matrix, \textit{LittleAriadne} will return, by default, top 40 most related entities, which could be topical terms, authors, subjects, citations or clusters. If there are multiple known entities in the search query, a weighted average of the vectors of individual entities is used to calculate similarities (the same way an article vector is constructed). If the search query does not contain any known entities, a blank page is returned, as there is no information about this query. 

Figure~\ref{fig.gammaray} gives a contextual view of ``gamma ray.''\footnote{Available at \url{http://thoth.pica.nl/astro/relate?input=gamma+ray}} The search query refers to an known topical term ``gamma ray,'' and it is therefore displayed as a red node in the network visualization. The top 40 most related entities are shown as nodes, with the top 5 connected by the red links. The different colours reflect their types, e.g., topical terms, subjects, authors, or clusters. Each of these 40 entities is further connected to its top 5 most related entities among the rest of the entities in the visualization, with the condition that the cosine similarity is not below 0.6. A thicker link means the two linked entities are \textit{mutually} related, i.e., they are among each other's top 5 list. The colour of the link takes that of the node where the link is originated. If the link is mutual and two linked entities are of different types, one of the entity colours is chosen.  

The displayed entities often automatically form groups depending on their relatedness to each other, whereby more related entities are positioned closer to each other. Each group potentially represents a different aspect related to the query term. The size of a node is proportional to the logarithm of its frequency of occurrences in the whole dataset. The absolute number of occurrences appears when hovering the mouse cursor over the node. Due to the fact that different statistical methods are at the core of the \textit{Ariadne} algorithm, this number gives an indication of the reliability of the suggested position and links.

In Figure~\ref{fig.gammaray}, there are four clusters from OCLC-31, ECOOM-BC13 and ECOOM-NLP11, and CWTS. The ECOOM-BC13 cluster $eb\ 8$ and ECOOM-NLP11 cluster $en\ 4$ are directly linked to ``gamma ray,'' suggesting that  these two clusters are probably about gamma rays. It is not surprising that they are very close to each other, because they contain 7560 and 5720 articles respectively but share 3603 articles. At the lower part, the OCLC-31 cluster $ok\ 21$ and the CWTS cluster $c\ 15$ are also pretty close to our search term. They contain 1849 and 3182 articles respectively and share 1721 articles in common which makes them close to each other in the visualization. By looking at the topical terms and subjects around these clusters, we can have a rough idea of their differences. Although they are all about ``gamma ray,'' Clusters $eb\ 8$ and $en\ 4$ are probably more about ``radiation mechanisms,'' ``very high energy,'' and ``observations,'' while Clusters $ok\ 21$ and $c\ 15$ seem to focus more on ``afterglows,''  ``prompt emission,'' and ``fireball.'' Such observations will invite users to explore these clusters or subjects further. 

\begin{figure}
\centering 
\includegraphics[width=\linewidth]{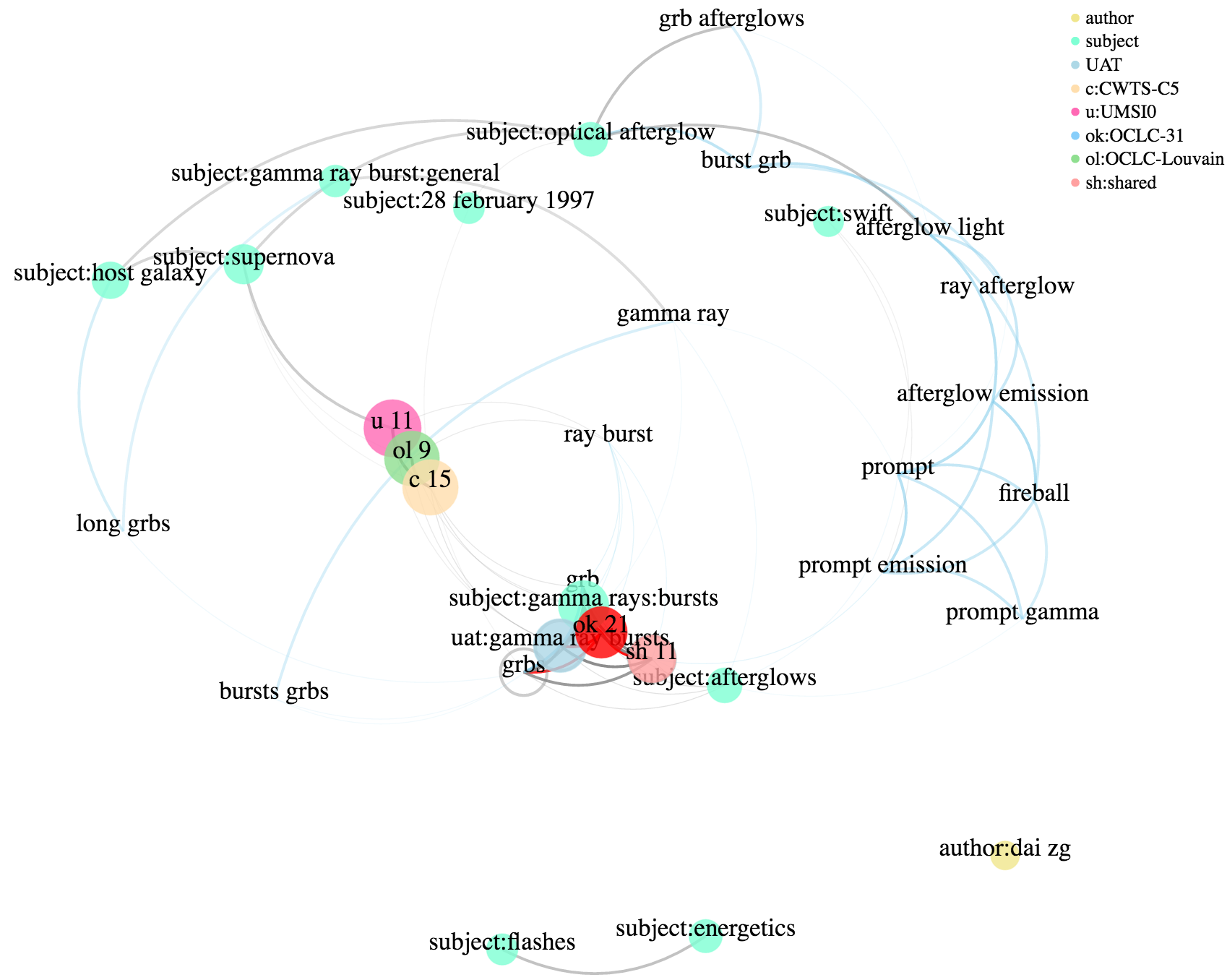}
\caption{The contextual view of cluster $ok\ 21$ \label{fig.ok21}}
\end{figure}

Each node is clickable which leads to another visualization of the context of this selected node. If one is interested in cluster $ok\ 21$ for instance, after clicking the node, a contextual view of cluster $ok\ 21$ is presented,\footnote{Available at \url{http://thoth.pica.nl/astro/relate?input=[cluster:ok\%2021]}. } as shown in Figure~\ref{fig.ok21}. This context view 
provides a good indication about the content of the articles grouped together in this cluster. In the context view of cluster $ok\ 21$ we see again the cluster $c\ 15$, which was already near to $ok\ 21$ in the context view of ``gamma ray.'' But the two ECOOM clusters, $eb\ 8$ and $en\ 4$ that are also in the context  of ``gamma ray'' are not visible any more. Instead, we find two more similar clusters $u\ 11$ and $ol\ 9$. That means that,  even though the clusters $ok\ 21$ and $eb\ 8$ are among the top 40 entities that are related to ``gamma ray,'' they are still different in terms of their content. This can be confirmed by looking at their labels in Table~\ref{tab.ok21}.\footnote{More details about cluster labelling can be found in~\cite{koopman2015_labelling}.}

\begin{table}[t]
\centering
\caption{Labels of clusters similar to $ok\ 21$ and to ''gamma ray''
\label{tab.ok21}}
{
\begin{tabularx}{\textwidth}{clX}\hline
Cluster IDs & Size & Cluster labels\\\hline
ok 21 & 	1849	& grb, ray burst, gamma ray, afterglow, bursts grbs, swift, prompt emission, prompt, fireball, batse\\
c 15	&	3182 &	grb, ray bursts, gamma ray, afterglow, bursts grbs, sn, explosion, swift, type ia, supernova sn\\
ol 9 & 	2895	& grb, ray bursts, gamma ray, afterglow, bursts grbs, sn, type ia, swift, explosion, ia supernovae\\
u 11 & 2051	& grb, ray bursts, gamma ray, afterglow, bursts grbs, sn, explosion, type ia, swift, supernova\\\hline
eb 8 & 7560	 & gamma ray, pulsar, ray bursts, grb, bursts grbs, high energy, jet, radio, psr, synchrotron\\
en 4 & 5720 & gamma ray, grb, ray bursts, cosmic ray, high energy, bursts grbs, afterglow, swift, tev, tev gamma
\\\hline
\end{tabularx}}
\end{table}



As mentioned before, in the interface one can also further refine the display. For instance, one can choose the number of nodes to be shown or decide to limit the display to only authors, journals, topical terms, subjects, citations or clusters. The former can be done by the slider $show$ or by editing  the URL string directly. For the latter options, tick boxes are given. An additional slider $font$ allows to experiment with the font size of the labels. 

\begin{figure}[t]
\includegraphics[width=\linewidth]{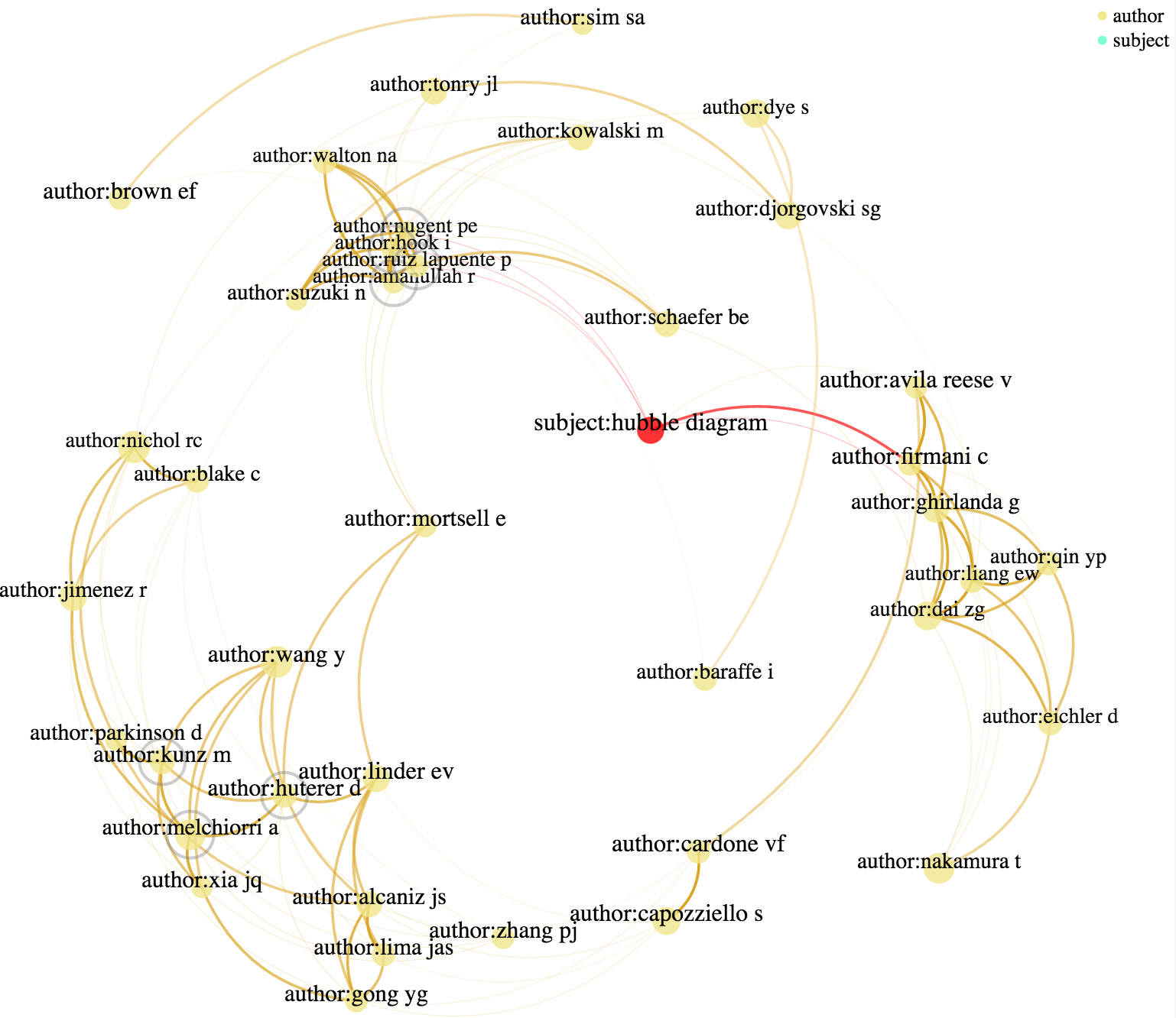}
\caption{The authors who are the most related to ``subject:hubble diagram" \label{fig.author}}
\end{figure}

A display with only one type of entity enables us to see context filtered along one perspective (lexical, journals, authors, subjects), and is often useful. For example, Figure~\ref{fig.author}\footnote{Available at \url{http://thoth.pica.nl/astro/relate?input=\%5Bsubject\%3Ahubble+diagram\%5D&type=2}} shows at least three separate groups of authors who are most related to ``subject:hubble diagram.'' 

At any point of exploration, one can see the most related entities, grouped by their types and listed at the bottom of the interface. The first category shown are the \textit{related titles}, the titles of the articles most relevant to a search query. Due to license restrictions, we cannot make the whole bibliography available. But when clicking on a title, one actually sees the context of a certain article. Not only titles can be clicked through, all entities at the lower part are also clickable and such an action leads to another contextual view of the selected entity.

At the top of the interface, under the search box, we find further hyperlinks behind the label \textit{exact search} and \textit{context search}. Clicking on the hyperlinks  automatically sends queries to other information spaces such as Google, Google Scholar, Wikipedia, and WorldCat. For \textit{exact search}, the same query text is used. For \textit{context search}, the system generates a selection among all topical terms related to the original query term and send this selection as a string of terms (with the Boolean AND operation) to those information spaces behind the hyperlinks. This option offers users a potential way to retrieve related literature or web resources from a broader perspective. In turn, it also enables the user to better understand the entity-based context view provided by \textit{Ariadne}.


Let us now come back to our first research question: how does the \textit{Ariadne} algorithm work on a much smaller, field-specific dataset? The interface shows that the original \textit{Ariadne} algorithm works well on the small \textit{Astro} dataset. Not surprisingly, compared with our exploration in the much bigger and more general ArticleFirst dataset, we find more consistent representations; that is, specific vocabulary is displayed, which can be cross-checked in Wikipedia, Google or Google Scholar. 
On the other hand, different corpora introduce different contexts for entities. For example, ``young'' in ArticleFirst\footnote{Available at \url{http://thoth.pica.nl/relate?input=young}} is associated with adults and 30 years old, while in \textit{LittleAriadne} it is immediately related to young stars which are merely 5 or 10 millions years old.\footnote{Available at \url{http://thoth.pica.nl/astro/relate?input=young}} Also, the bigger number of topical terms in the larger database leads to a situation where almost every query term produces a response. In \textit{LittleAriadne} searches for, e.g., a writer such as $Jane Austen$ retrieve nothing. Not surprisingly, for domain-specific entities, \textit{LittleAriadne} tends to provide more accurate context. A more thorough evaluation needs to be based, as for any other topical mapping, on a discussion with domain experts.   



\subsection{Experiment 2 -- Comparing clustering solutions}

In \textit{LittleAriadne} we extended the interface with the goal of observing and comparing clustering solutions visually. As discussed in Section~\ref{sec.rp} cluster assignments are treated in the same way as other entities associated with articles, such as topical terms, authors, etc. Each cluster ID is therefore represented in the same  space and visualized in the same way. In the interface, when we use a search term, for example ``[cluster:c]'' and tick the ``scan'' option, the interface scans all the entities in the semantic matrix which starts with, in this case ``cluster:c,'' and then effectively selects and visualizes all CWTS-C5 clusters.\footnote{This \textit{scan} option is applicable to any other type of entities, for example, to see all subjects which start with ``quantum'' by using ``subject:quantum'' as the search term and do the scanning.} This way, we can easily see the distribution of a single clustering solution. Note that in this scanning visualization, any cluster which contains less than 100 articles is not shown. 

\begin{figure}
\centering
\begin{tabular}{lcr}\\
\includegraphics[width=0.4\linewidth]{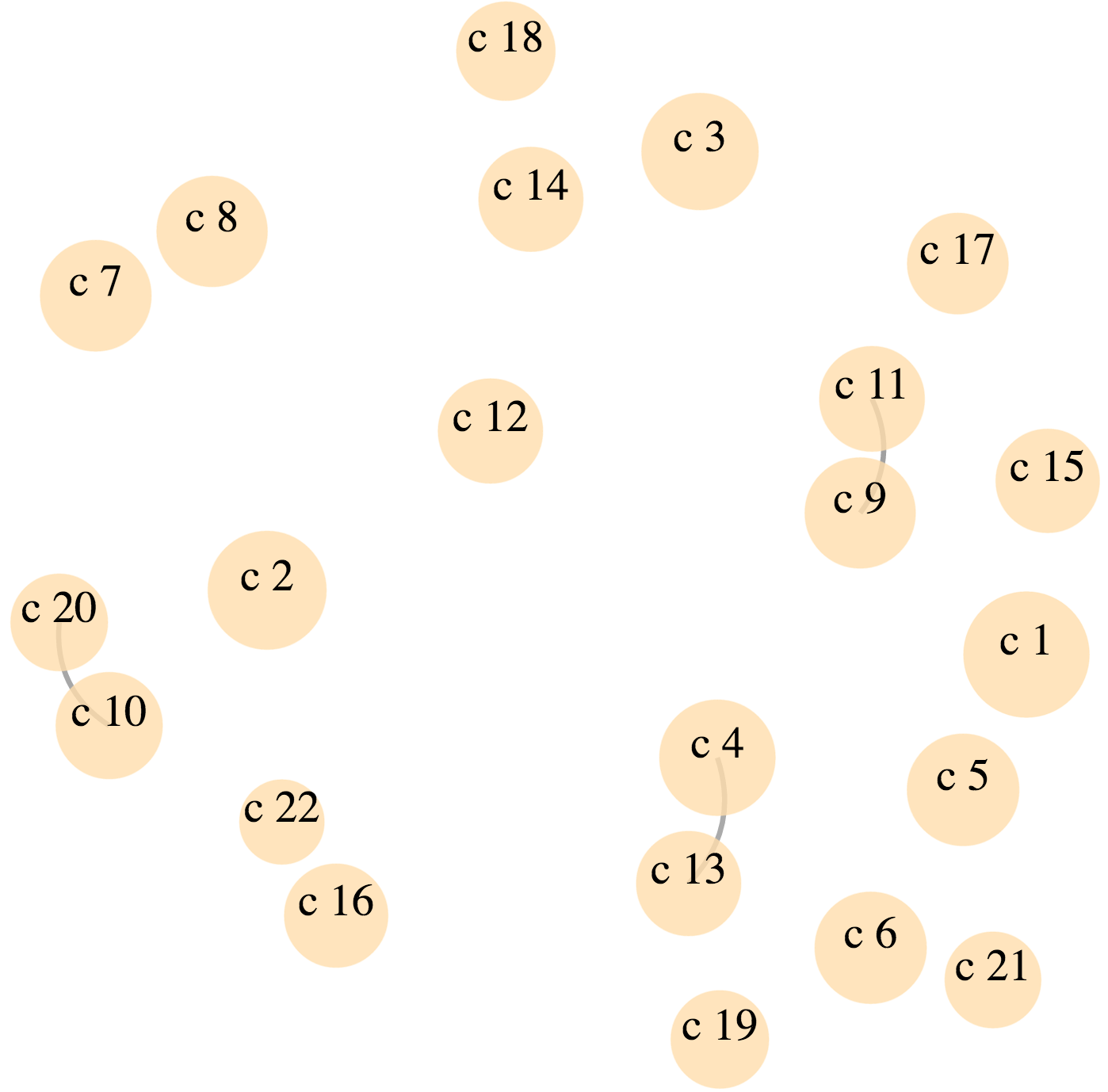} && \includegraphics[width=0.4\linewidth]{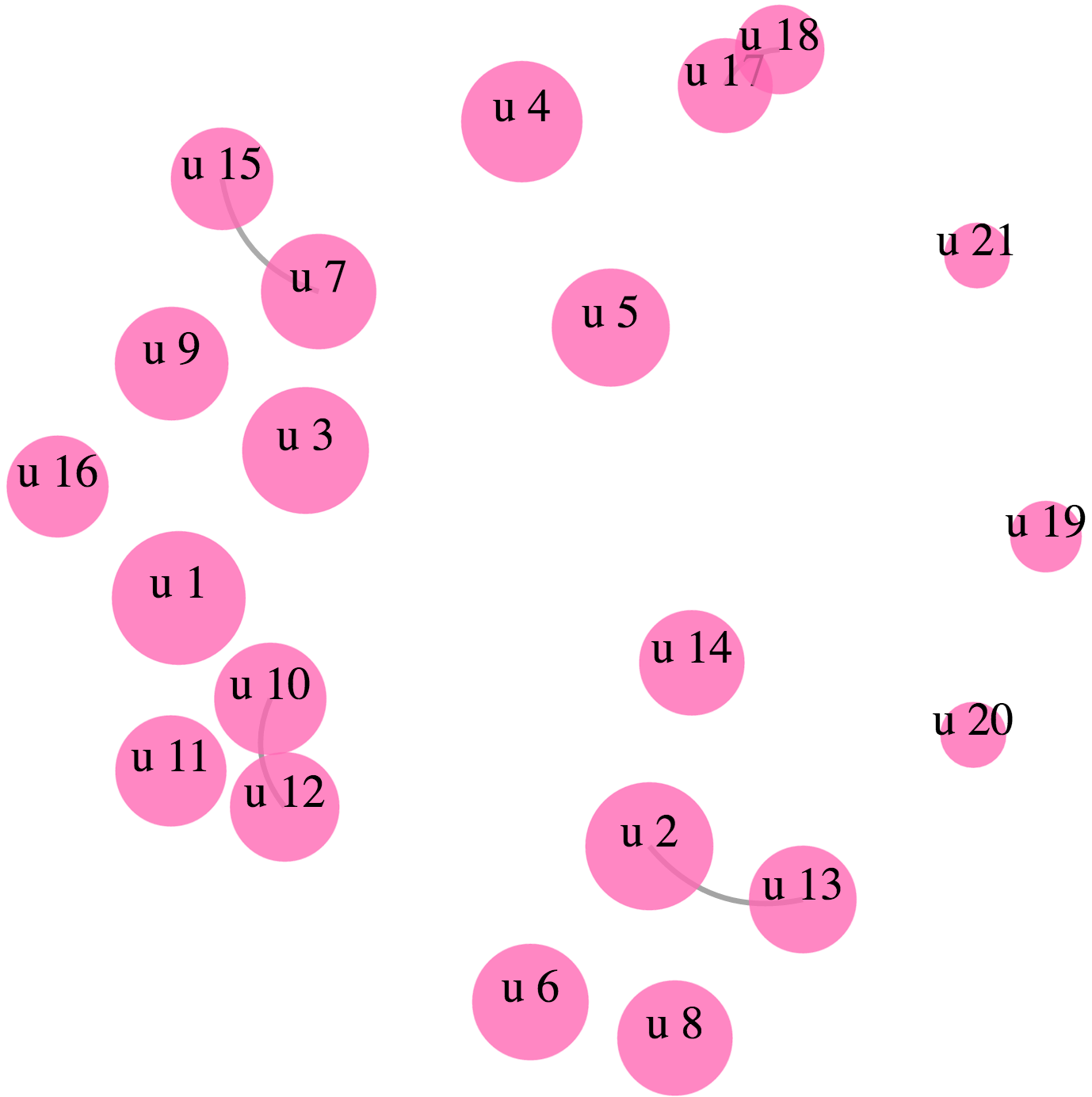}\\
(a) CWTS-C5 clusters && (b) UMSI0 clusters\\ 
\includegraphics[width=0.4\linewidth]{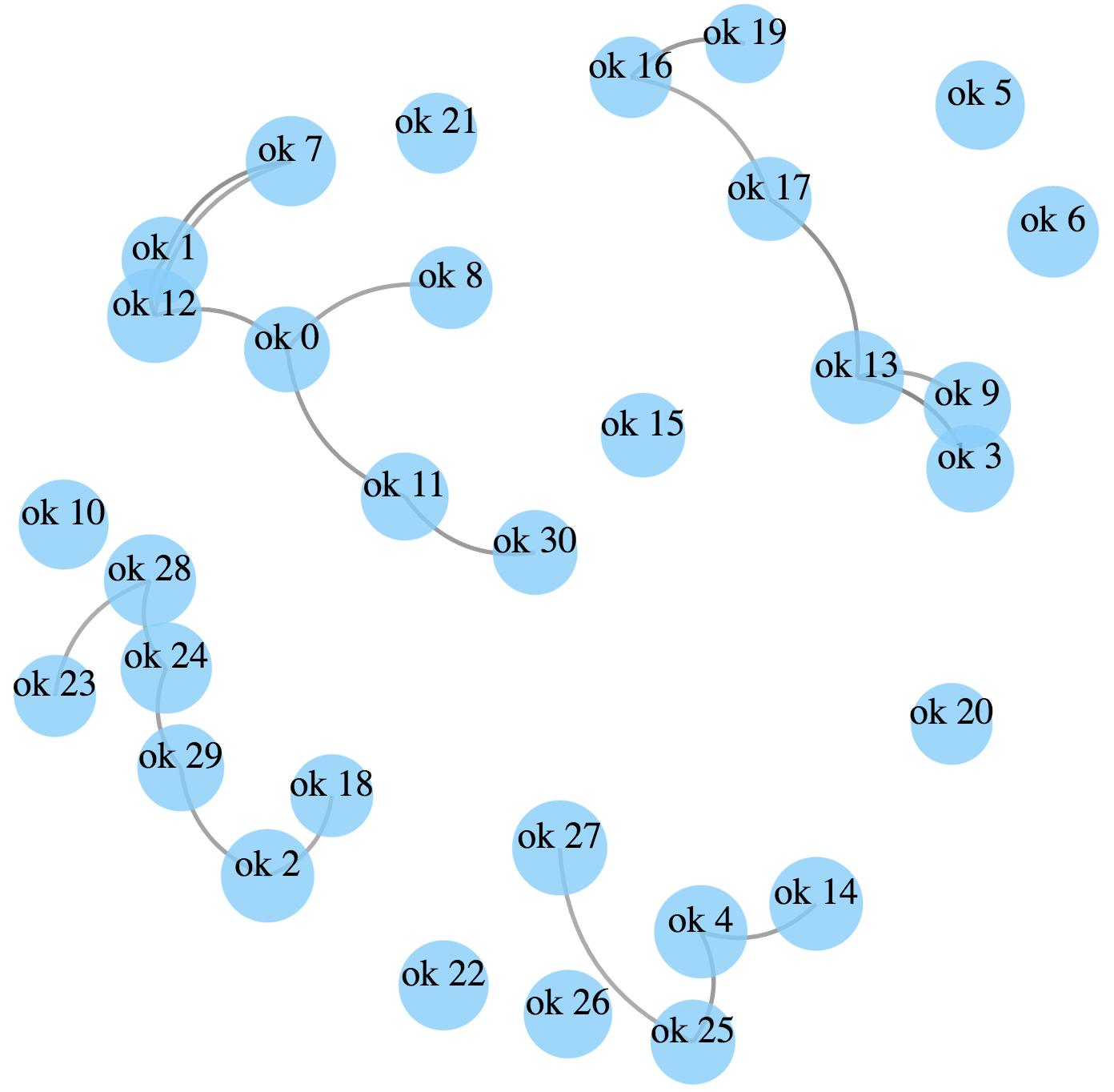} && \includegraphics[width=0.4\linewidth]{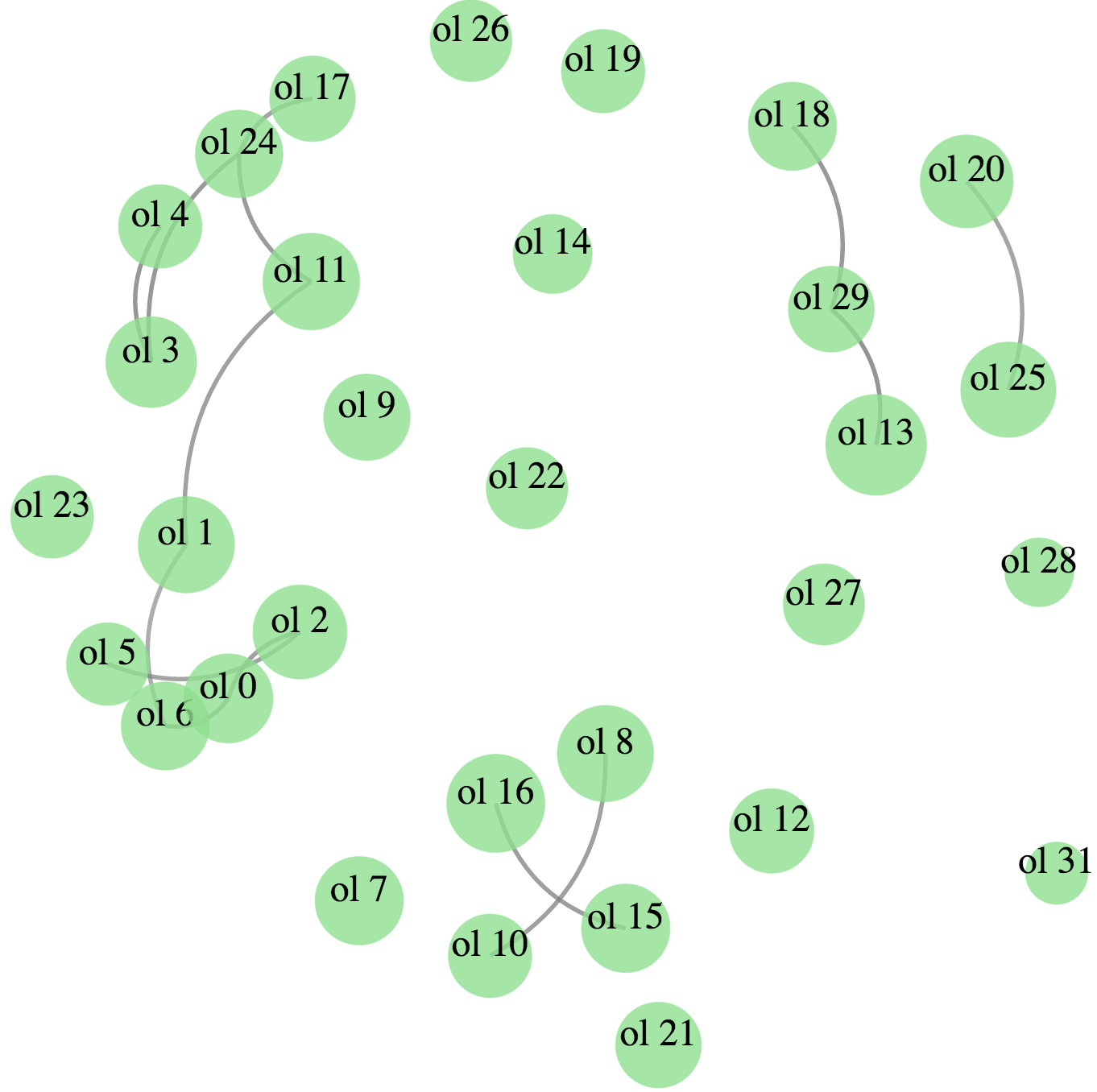}\\
(c) OCLC-31 clusters && (d) OCLC-Louvain clusters \\
\includegraphics[width=0.4\linewidth]{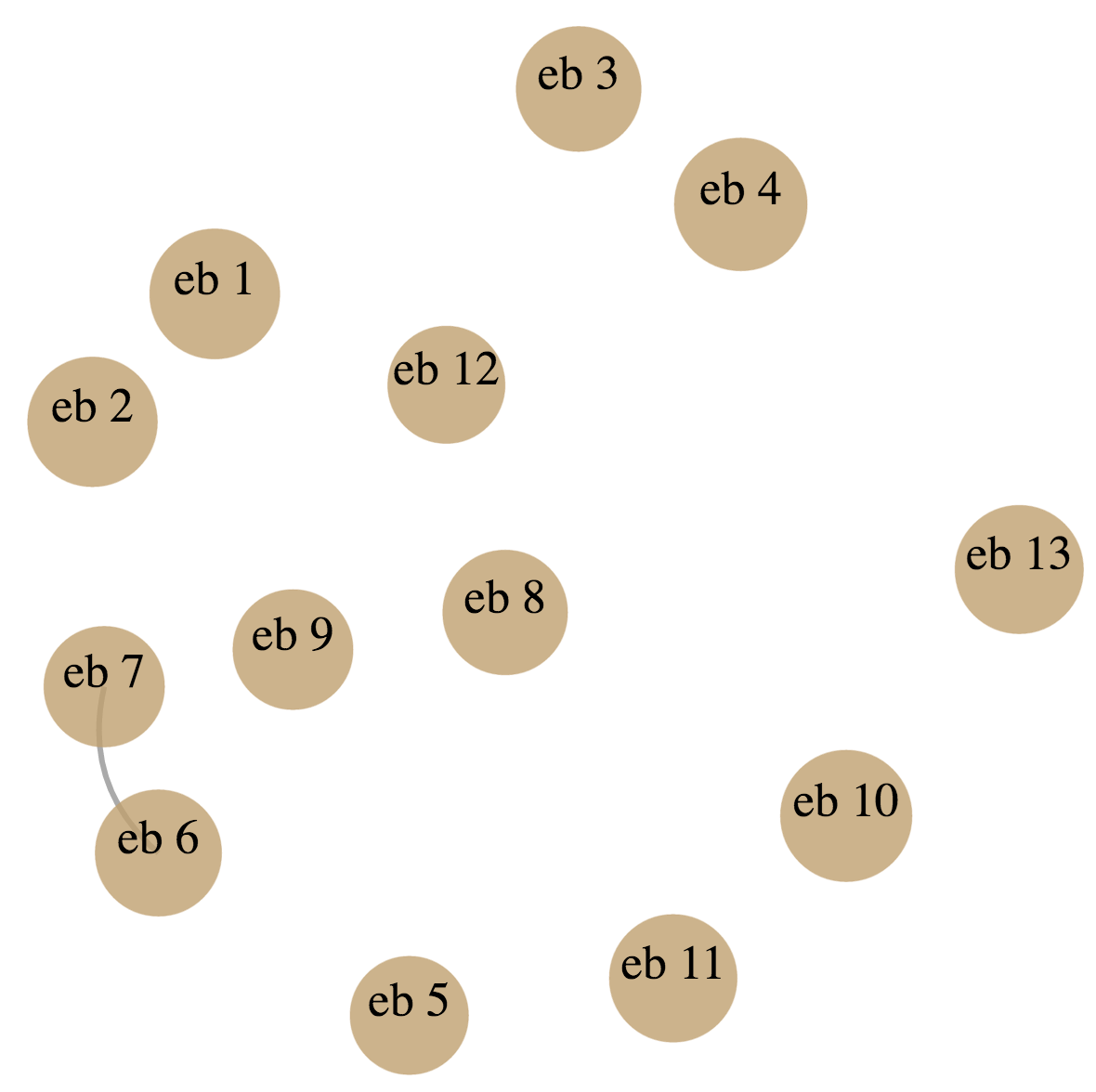} && \includegraphics[width=0.4\linewidth]{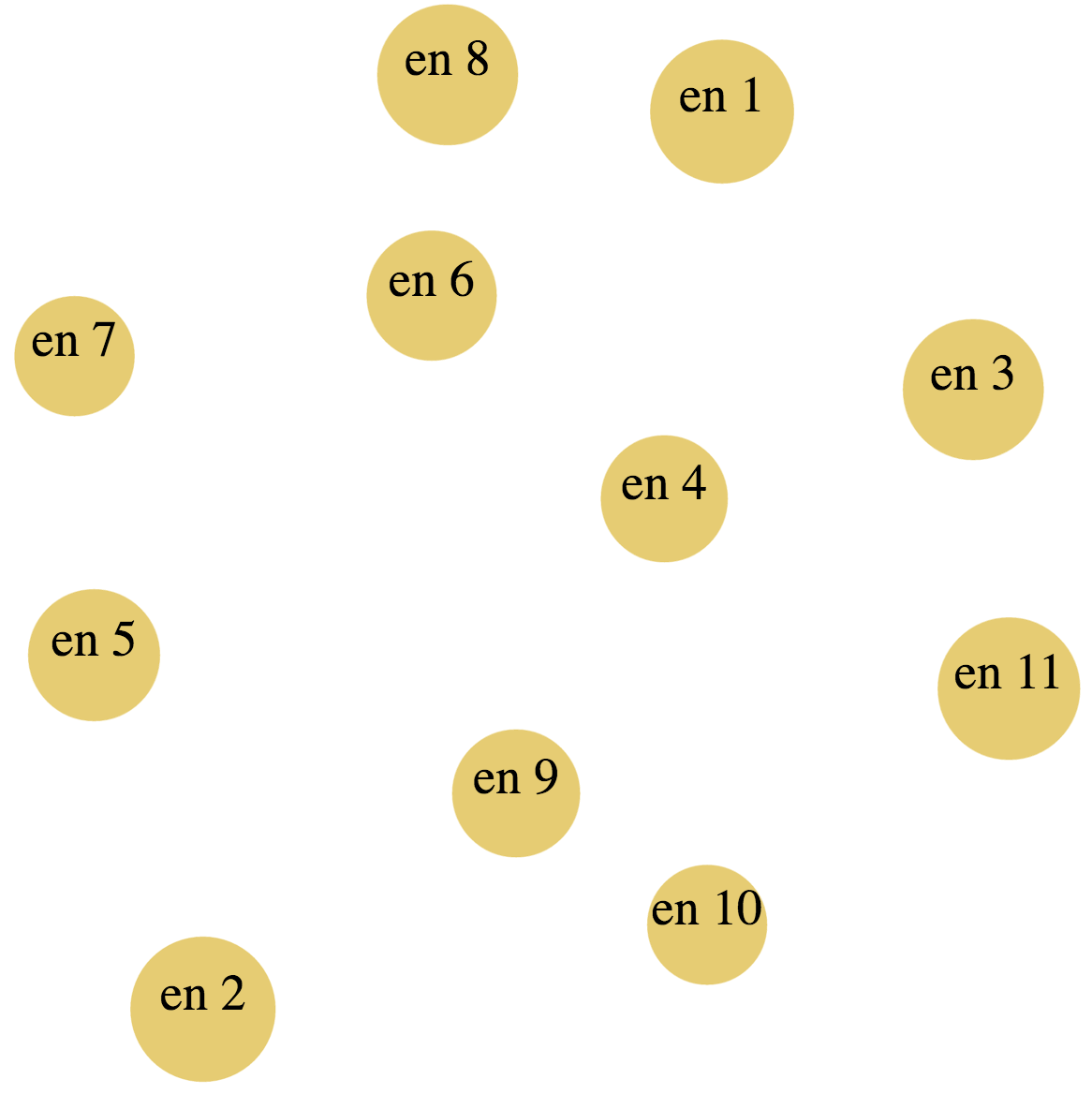}\\
(e) ECOOM-BC13 clusters && (f) ECOOM-NLP11 clusters \\
\includegraphics[width=0.4\linewidth]{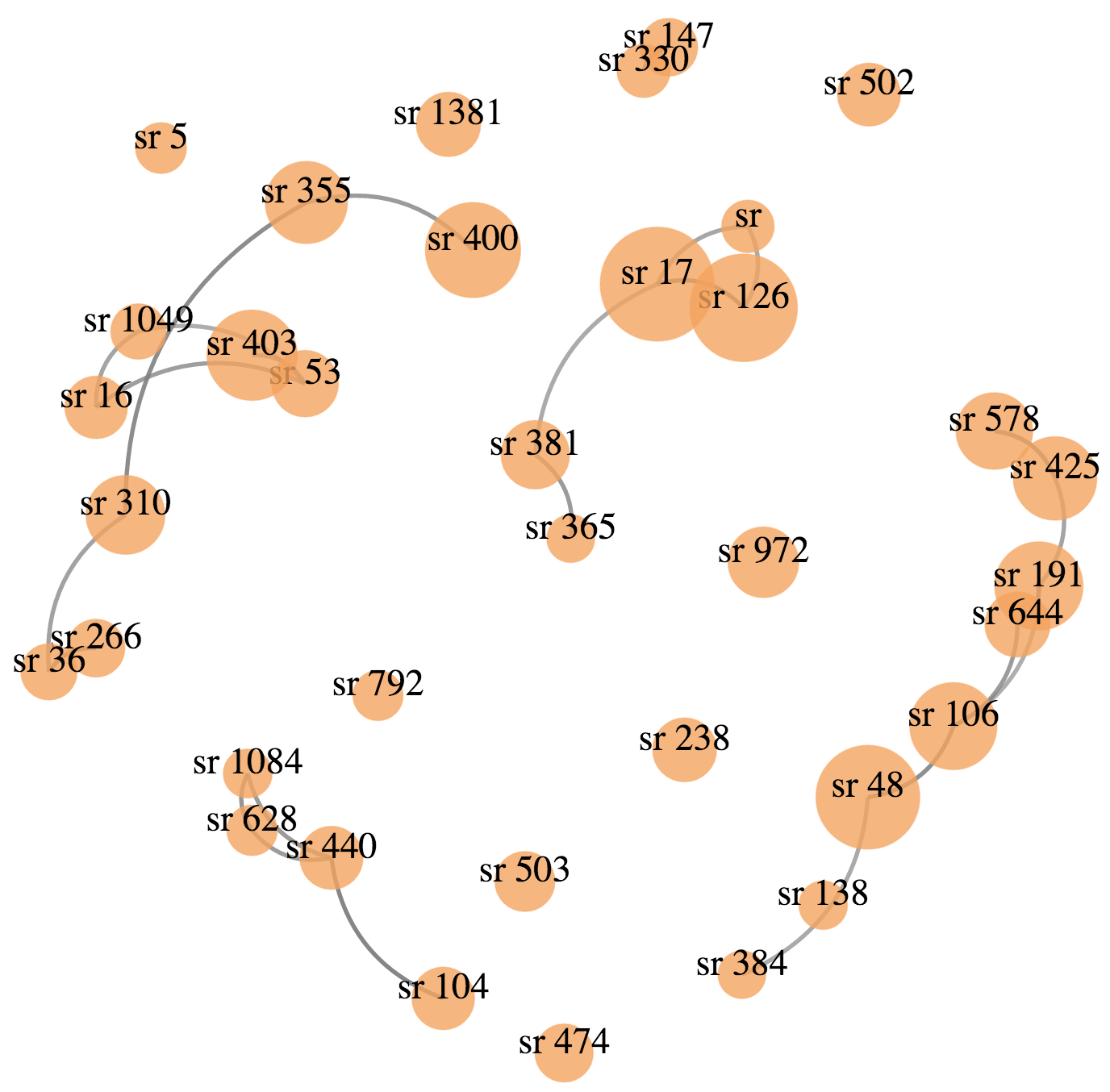} && \includegraphics[width=0.4\linewidth]{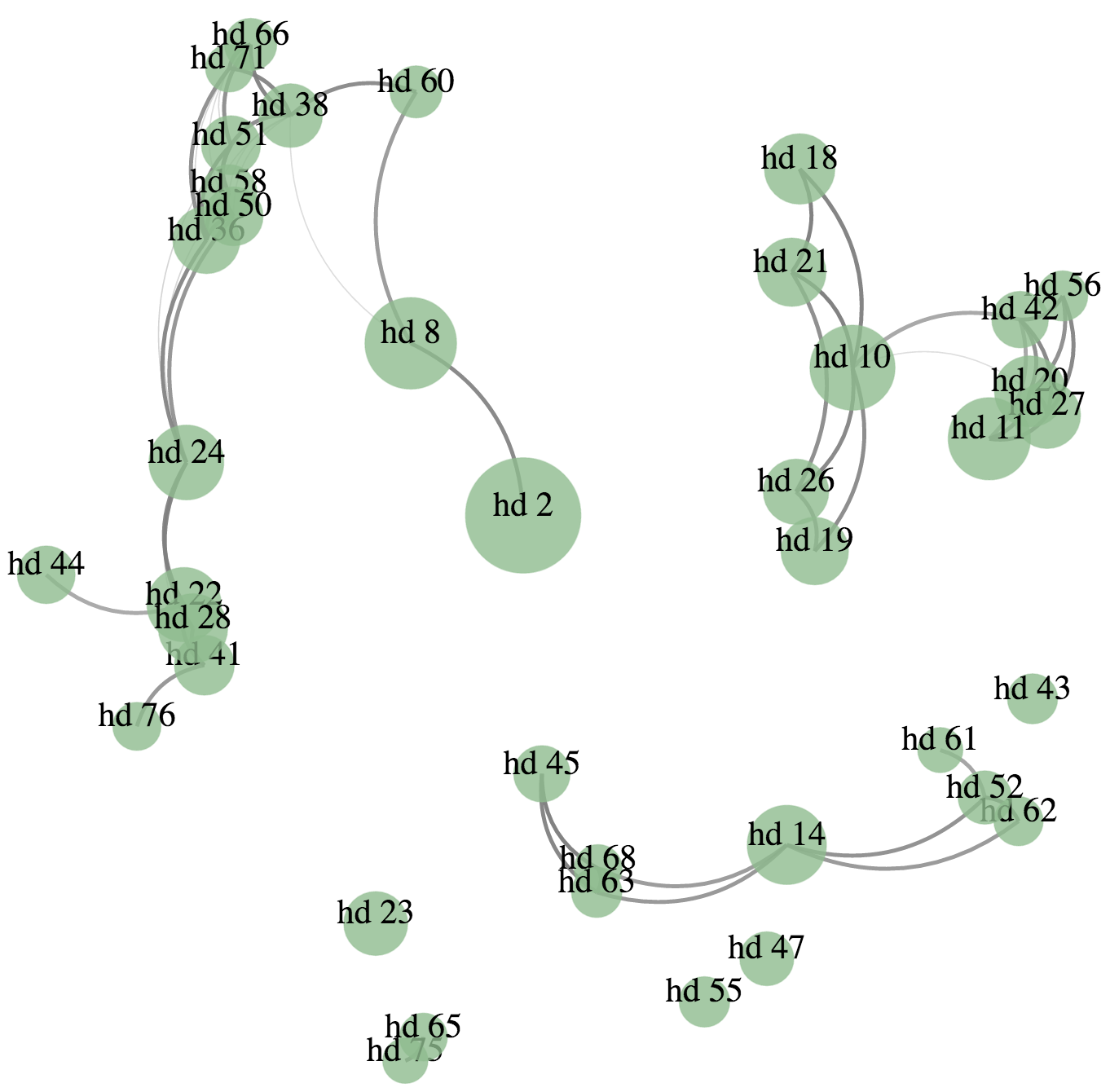}\\
(g) STS-RG clusters && (h) HU-DC clusters \\
\end{tabular}
\caption{The distribution of clusters \label{fig.distribution}}
\end{figure}

Figure~\ref{fig.distribution} shows the individual distribution of clusters from all eight clustering solutions. When two clusters have a relatively high mutual similarity, there is a link between them. It is not surprising to see the HU-DC clusters are highly connected as they are overlapping, and form a poly-hierarchy. Compared to CWTS-C5, UMSI and two ECOOM clusters, the STS-RG and the two OCLC solutions have more cluster-cluster links. This suggests that these clusters overlap more in terms of their direct vocabularies and indirect vocabularies associated with their authors, journals and citations.

\begin{figure}
\centering
\begin{tabular}{c}
\includegraphics[width=.8\linewidth]{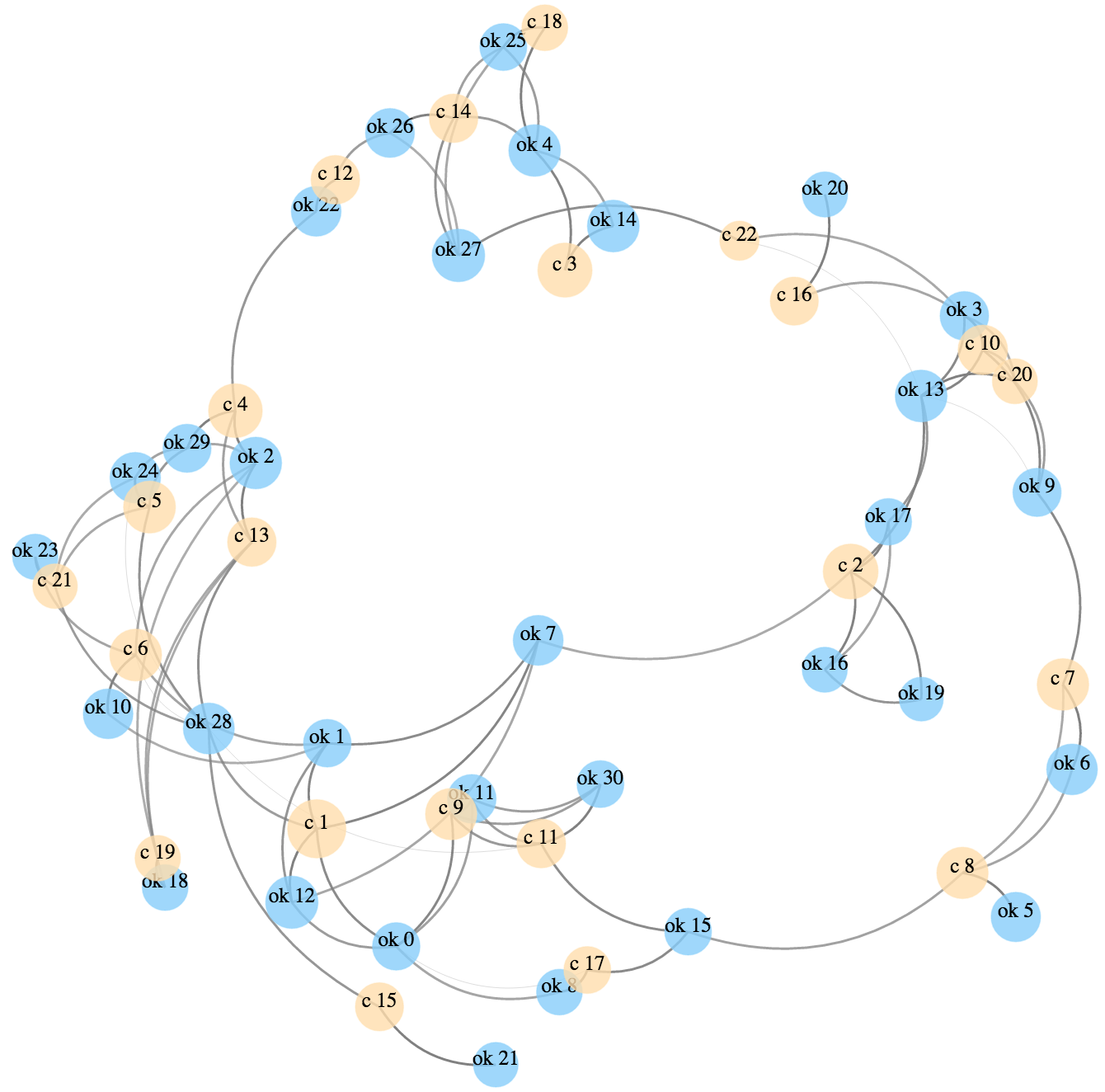}\\
(a) Highly similar clustering solutions\\
\includegraphics[width=.8\linewidth]{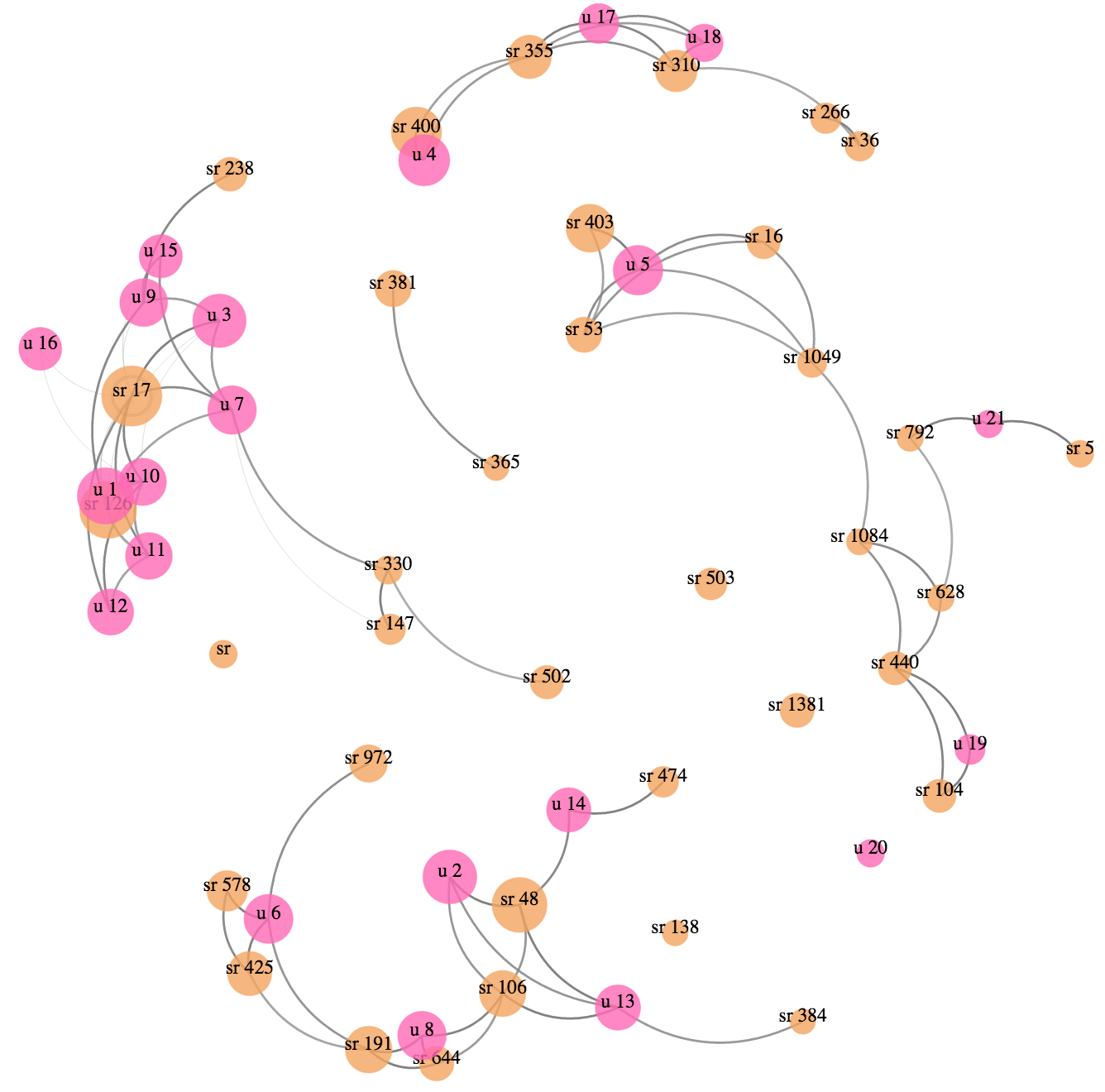}\\
 (b) Clustering solutions with different focuses\\
\end{tabular}
\caption{Visual comparison of clustering solutions \label{fig.compare}}
\end{figure}
\begin{figure}
\centering
\begin{tabular}{c}
\includegraphics[width=.8\linewidth]{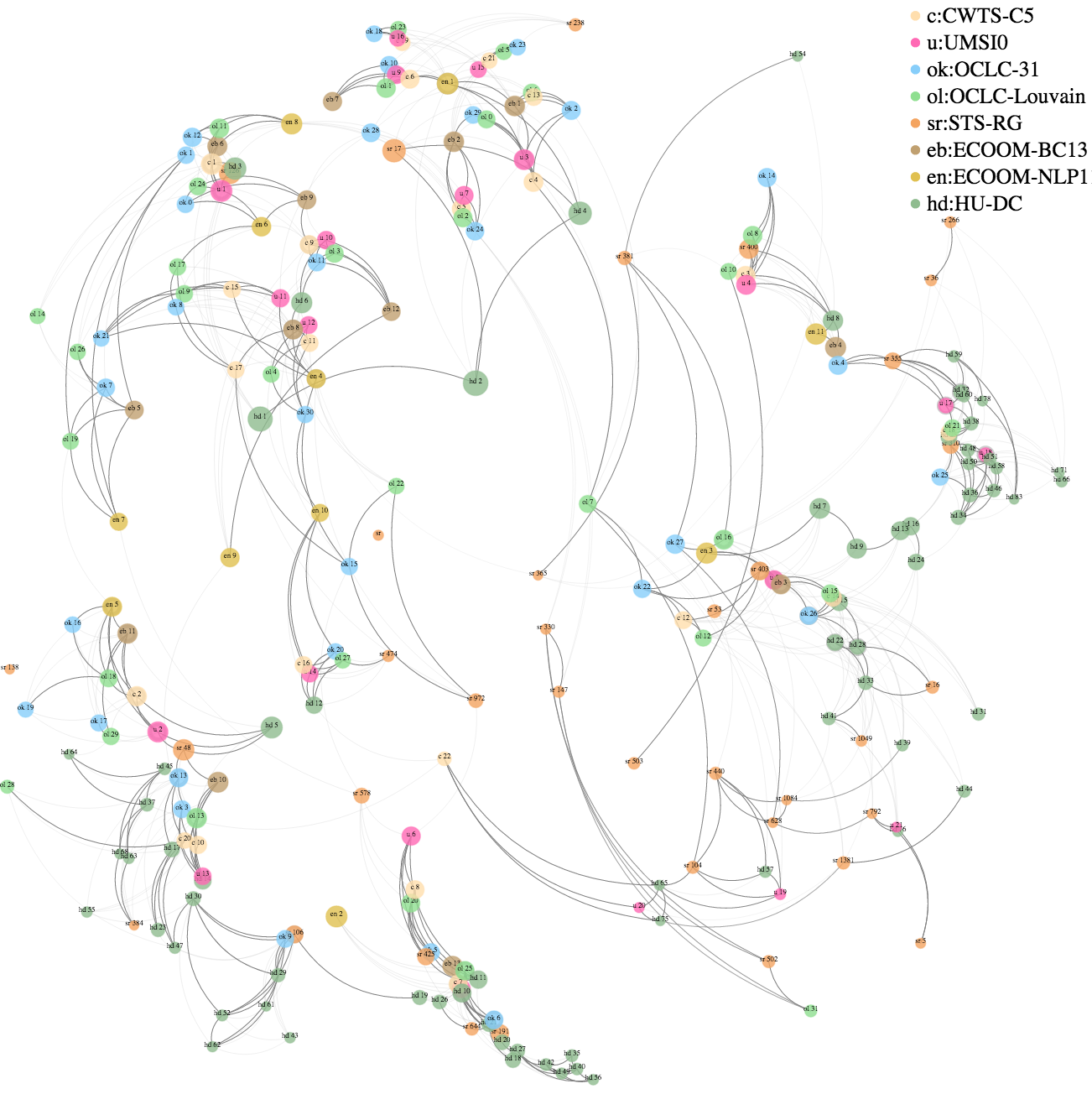}\\
(a) All clustering solutions\\
\includegraphics[width=.8\linewidth]{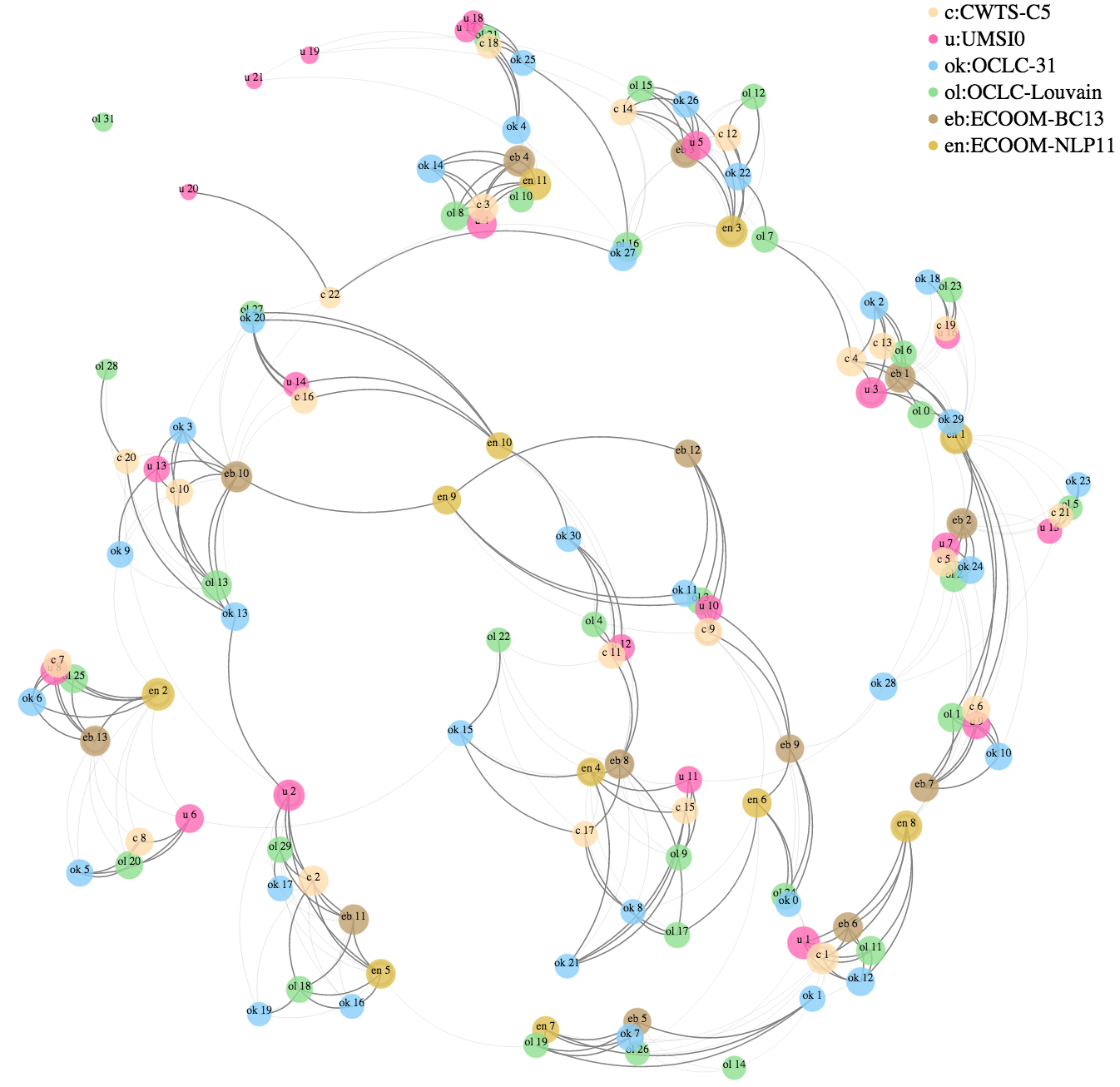}\\
 (b) Clusters from CWTS, UMSI, OCLC and ECOOM\\
\end{tabular}
\caption{Visual comparison of clustering solutions\label{fig.allclusters}}
\end{figure}

If we scan two or more cluster entities, such as ``[cluster:c][cluster:ok],'' we put two clustering solutions on the same visualization so that they can be compared visually. In Figure~\ref{fig.compare} (a) we see the high similarity between clusters from CWTS-C5 and those from OCLC-31.\footnote{Available at \url{http://thoth.pica.nl/astro/relate?input=\%5Bcluster\%3Ac\%5D\%5Bcluster\%3Aok\%5D&type=S&show=500}} CWTS-C5 has 22 clusters while OCLC-31 has 31 clusters. Each CWTS-C5 cluster is accompanied by one or more OCLC clusters. This indicates that they are different, probably because of the granularity aspect instead of any fundamental issue. Figure~\ref{fig.compare} (b) shows two other sets of clusters that partially agree with each other but clearly have different capacity in identifying different clusters.\footnote{Available at \url{http://thoth.pica.nl/astro/relate?input=\%5Bcluster\%3Au\%5D\%5Bcluster\%3Asr\%5D&type=S&show=500}} 

Figure~\ref{fig.allclusters} (a) shows all the cluster entities from all eight clustering solutions.\footnote{Available at \url{http://thoth.pica.nl/astro/relate?input=\%5Bcluster\%3Ac\%5D\%5Bcluster\%3Au\%5D\%5Bcluster\%3Aok\%5D\%5Bcluster\%3Aol\%5D\%5Bcluster\%3Aeb\%5D\%5Bcluster\%3Aen\%5D\%5Bcluster\%3Asr\%5D\%5Bcluster\%3Ahd\%5D&type=S&show=500}} The STS and HU have hundreds of clusters, which make the visualization pretty cluttered.  Figure~\ref{fig.allclusters} (b) shows only the solutions from CWTS, UMSI, OCLC and ECOOM, whose numbers of the clusters are comparable.\footnote{Available at \url{http://thoth.pica.nl/astro/relate?input=\%5Bcluster\%3Ac\%5D\%5Bcluster\%3Au\%5D\%5Bcluster\%3Aok\%5D\%5Bcluster\%3Aol\%5D\%5Bcluster\%3Aeb\%5D\%5Bcluster\%3Aen\%5D&type=S&show=500}}

Concerning our second research question - can we use \textit{LittleAriadne} to compare clustering solutions visually? - we can give a positive answer. But, it is not easy to see from \textit{LittleAriadne} why some clusters are similar and the others not. The visualization functions as a \textit{macroscope}\cite{boerner2011} and provides a general overview of all the clustering solutions, which helps to guide further investigation. It is not conclusive, but a useful heuristic devise. For example, from Figure~\ref{fig.allclusters}, especially~\ref{fig.allclusters} (b), it is clear that there are ``clusters of clusters.'' That is, some clusters are detected by all of these different methods. In the future we may investigate these clusters of clusters more closely and perhaps discover that different solutions identify some of the same topics. We continue the discussion of the use of visual analytics to compare clustering solutions in the paper by Velden et al. \cite{velden2015comparison}.

\section{Conclusion}
\label{sec.conclusion}
We present a method implemented in an interface that allows browsing through the context of entities, such as topical terms, authors, journals, subjects and citations associated with a set of articles. With the \textit{LittleAriadne} interface, one can navigate visually and interactively through the context of entities in the dataset by seamlessly travelling between authors, journals, topical terms, subjects, citations and cluster IDs as well as consult external open information spaces for further contextualization. 

In this paper we particularly explored the usefulness of the method to the problem of topic delineation addressed in this special issue. \textit{LittleAriadne} treats cluster assignments from different solutions as additional special entities. This way we provide the contextual view of clusters as well. This is beneficial for users who are interested in travelling seamlessly between different types of entities and their related cluster assignments generated by different solutions. 

We also contributed two clustering solutions built on the vector representation of articles, which is  different from solutions provided by other methods. We start by including references and treating them as entities with a certain lexical or semantic profile. In essence, we start from a multipartite network of papers, cited sources, terms, authors subjects, etc. and focus on similarity in a high dimensional  space.
Our clusters are comparable to other solutions yet have their own characteristics. Please see~\cite{velden2015comparison,koopman2015_clustering} for more details. 

We demonstrated that we can use \textit{LittleAriadne} to compare different clustering solutions visually and  generate a wider overview. This has a potential to be complementary to any other method of cluster comparison. We hope that this interactive tool supports discussion about different clustering algorithms and helps to find the right meaning of clusters. 

We have plans to further develop the \textit{Ariadne} algorithm. The \textit{Ariadne} algorithm is general enough to incorporate additional types of entities into the semantic matrix. Which entities we can add very much depends on the information in the original dataset or database. In the future, we plan to add  publishers, conferences, etc. with the aim to provide a richer contextualization of entities typically found in a scholarly publication. We also plan to elaborate links to articles that contribute to the contextual visualization, thus strengthening the usefulness of \textit{Ariadne} not only for the associative exploration of contexts similar to scrolling through a systematic catalogue, but also as a direct tool for document retrieval.  

In this context we plan to further compare \textit{LittleAriadne} and \textit{Ariadne}. As mentioned before, the corpora matter when talking about context of entities. The advantage of \textit{LittleAriadne} is the confinement of the dataset to one scientific discipline or field and topics within. We hope by continuing such experiments also to learn more about the relationship between genericity and specificity of contexts, and how that can be best addressed in information retrieval.

\section*{Acknowledgement}
Part of this work has been funded by the COST Action TD1210 Knowescape, and the FP7 Project ImpactEV. We would like to thank the internal reviewers Frank Havemann, Bart Thijs as well as the anonymous external referees for their valuable comments and suggestions. We would also like to thank Jochen Gl\"aser, William Harvey and Jean Godby for comments on the text. 

\bibliographystyle{spmpsci}      
\bibliography{astro}   

\end{document}